# Tunneling characteristics of graphene

Young Jun Shin,[1,2] Gopinadhan Kalon,[1,2] Jaesung Son,[1] Jae Hyun Kwon,[1,2] Jing Niu,[1] Charanjit S. Bhatia,[1] Gengchiau Liang,[1] and Hyunsoo Yang [1,2,a)]

[1]*Department of Electrical and Computer Engineering, National University of Singapore, Singapore*

[2]*NUSNNI-NanoCore, National University of Singapore, Singapore*

Negative differential conductance and tunneling characteristics of two-terminal graphene devices are observed before and after electric breakdown, respectively. The former is caused by the strong scattering under a high E-field, and the latter is due to the appearance of a tunneling barrier in graphene channel induced by a structural transformation from crystalline graphene to disordered graphene because of the breakdown. Using Raman spectroscopy and imaging, the presence of non-uniform disordered graphene is confirmed. A memory switching effect of 100000% ON/OFF ratio is demonstrated in the tunneling regime which can be employed in various applications.

[a)] e-mail address: eleyang@nus.edu.sg

Graphene, a single layer of carbon atoms, has attracted tremendous attention in the last few years due to its potential applications in the area of ultrafast devices.[1, 2] The successful demonstration of 100 GHz graphene transistor and the success of various methods to produce large scale graphene are indicating that graphene can be a promising material in the post-silicon era.[3, 4] Although graphene has superior physical properties such as very high mobility (Fermi velocity $v_F$~$10^6$ m/s), a low ON/OFF ratio due to its semi-metallic property prevents graphene from being engineered as logic and memory devices. Two-terminal devices coupled with a nonlinear current-voltage (*I-V*) characteristic can be a solution to overcome this shortcoming.[5-7]

In this Letter, the tunneling behavior in graphene at a high bias voltage is reported. A high voltage bias is applied across the electrodes to induce a breakdown in the graphene channel. Negative differential conductance is found just before the breakdown as a symptom of the tunneling behavior. Understanding the behavior of carrier transport in graphene under a high bias voltage is important since graphene holds promising future as various components including interconnects because of its high current carrying capacity. After the breakdown, non-linear *I-V* curves, which have the same characteristic as tunneling diodes, are measured. Graphene is examined by Raman spectroscopy and it appears that crystal graphene is transformed into non-uniform disordered graphene under the application of high bias. Reproducible tunneling behavior and current hysteresis with a 100000% ON/OFF ratio are observed after the breakdown. The tunneling effect can be attributed to a non-uniform disordered system which introduces energy barriers in the graphene channel. To confirm this further, glassy carbon is uniformly deposited by pulsed laser deposition (PLD) and is subjected to the similar high bias voltage. The linear *I-V* characteristics of glassy carbon under a high bias condition indicate that the tunneling diode effect is a unique property of non-uniform disordered graphene.

Mechanical exfoliation is used to prepare single and multi-layer graphene.[8] Mechanically cleaved graphene is transferred to a highly p-doped Si substrate covered by a 300nm thick $SiO_2$ layer. Number of layers and quality of graphene is determined by Raman spectroscopy. Further details on the sample preparation and identification can be found in our earlier work.[9] Electrodes are patterned by standard lithography and Cr/Au (5 nm/80 nm) is deposited by a thermal evaporator. The contact deposition is followed by standard lift-off procedures. The *I-V* measurements with a two-probe configuration are carried out with a closed cycle helium cryostat under a base pressure of less than $1 \times 10^{-8}$ Torr at 3.8 K.

Single and multi-layer graphene samples are identified by Raman spectra as shown in Fig. 1(a) and the transport data from single layer graphene are reported in this work. In-plane vibrational *G* band (1580 cm$^{-1}$) and two-phonon 2*D* band (2670 cm$^{-1}$) are clearly visible without any indication of disorder *D* band peak and the number of layers of the graphene sample is determined by estimating the width of 2*D* peak.[10] Figure 1(b) shows the two-terminal resistance as a function of back gate voltage. From the shift of the Dirac point to the positive side of the applied back gate voltage, graphene is identified to be hole dominant. The Dirac point is shifted due to unintentional doping such as the adsorption of water molecules making graphene p-type.[8] The inset of Fig. 1(b) shows the optical image of a graphene device on top of a Si wafer with 300nm thick $SiO_2$ after the Cr/Au electrode deposition. The inset of Fig. 1(c) shows the typical linear *I-V* curve with the low bias voltage less than 2 V in graphene devices. As shown in Fig. 1(c), the *I-V* curve is linear at a low bias voltage, but becomes non-linear and slightly hysteretic at a higher bias.[11] Also the current starts to decrease as the applied voltage is continuously increased above 6 V. When the current reaches a maximum ($dI/dV = 0$), the current density is ~$1.9\times10^8$ A/cm$^2$ with a sample width of 16 µm and a thickness of 0.35 nm. When the *I-V* curve is numerically differentiated, we can clearly identify a negative differential conductance behavior as shown in Fig. 1(d). This negative differential conductance might be attributed to the self-heating effects caused by strong electron scattering due to hot non-equilibrium optical phonons similar to what is observed in carbon nanotubes.[12]

After the observation of negative differential conductance, the sweep voltage is increased further to induce a breakdown in graphene channel as was reported recently that graphene nanoribbons exhibit a breakdown current density, on the order of $10^8$ A/cm$^2$.[13] As can be seen in Fig. 2(a), the voltage sweep is halted as soon as the breakdown occurs. Figure 2(b) shows non-linear *I-V* behavior of graphene devices instead of linear increase in current with increasing the bias voltage, when graphene undergoes the breakdown. It can be clearly seen in the logarithmic scale that current starts increasing drastically when the bias exceed the threshold voltage of 7 V as shown in Fig. 2(c). After the measurement, graphene is examined by Raman spectroscopy and we find out that crystalline graphene is converted to disordered graphene. Interestingly, this disordered system is very different from typical disordered graphene created by an oxygen plasma treatment. Compared to the typical disordered system, the transformed graphene by an electrical breakdown has mixed phases such as amorphous-like (the upper inset of Fig. 2(a)) and graphene-like (the lower inset of Fig. 2(a)) phases. From the investigation by Raman spectroscopy, the breakdown can be considered as a "deforming process", which is the opposite concept of the forming process, the change in the spatial distribution of oxygen ions by applying the bias voltage, in the oxide resistance switching devices.[14] The deforming process introduces randomly distributed energy barriers caused by amorphous-like phase in the graphene channel.

A Raman imaging system is employed to investigate the details of the mixed phase graphene channel. The Raman images are plotted as the intensity of *D* and 2*D* band clearly show amorphous-like phase across the channel which is introduced by the breakdown. Since the deformed channel does not have a continuous current path after the breakdown, the charge carriers of graphene should go over the energy barrier due to the amorphous phase regime. As shown in Fig. 2(b) resistance is very high ($5.15\times10^{10}$ $\Omega$ at 3 V) below 7 V because the charge carriers have to tunnel through the barrier. On the other hand, resistance dramatically decreases to a lower value ($7.45\times10^{7}$ $\Omega$ at 15 V) under a high bias because the decrease of the effective tunneling width of the barrier caused by the electric field as sketched in the inset of Fig. 3(a). As a result, more carriers will tunnel through the barrier resulting higher currents. To strengthen this hypothesis, the *I-V* characteristics of a one-dimensional single-square barrier between two metal contacts are simulated based on non-equilibrium Green's function approach. The barrier height and width are set to 6 eV as the Fermi level ($E_F$) of graphene with respective to the vacuum level, and 1 nm, respectively. We find that its *I-V* characteristics in Fig. 3(a) are similar to the experimental results, and it indicates that the tunneling effect contributes the diode-like *I-V* characteristics.

The reproducibility of the tunneling diode effect is tested in the other sample by sweeping voltage several times from -6 to 6 V. It is reproducible with slight degradation as shown in Fig. 3(b). Degradation in the tunneling behavior is due to the enlargement of the barrier width caused by Joule heating. More than 20 devices are tested and it turns out that each device has a different threshold voltage since the breakdown is a random process. When the bias voltage is swept for even higher bias in the tunneling regime, current hysteresis is observed as plotted in Fig. 3(c). The range of the ON/OFF ratio is from 1000% to 100000%. Current hysteresis is repeatable but degraded gradually. To understand better and engineer the current hysteresis from the mixed phase channel, further studies are required.

In order to verify the unique nature of disordered graphene by the breakdown, we further study the *I-V* characteristics from other carbon thin films. For this a 2 nm thick glassy carbon film is deposited by pulsed laser deposition and tested in the same measurement conditions. From the inset of Fig. 3(d), the structure of deposited glassy carbon, strong *D* peak and weak 2*D* peak in Raman spectra, is similar to the structure of graphene after the breakdown. However, a complete mapping by Raman spectroscopy reveals that the glassy structure is very uniform unlike to the disordered graphene by the breakdown. When a high bias is applied across the glassy carbon channel, linear *I-V* characteristics are observed in Fig. 3(d), which demonstrates that the tunneling characteristic is a unique property of non-uniform disordered graphene. If the tunneling behavior in resistance of Fig. 2(b) is caused by electrical annealing, it should not be reversible and reproducible.[15] Non-linear *I-V* characteristics of graphene and carbon nanotubes due to a mechanical discontinuity have been reported.[5, 16, 17] We carefully check all our devices by scanning electron microscopy after the breakdown, but any mechanical discontinuity is not found as can be seen in the inset of Fig. 2(c).

In conclusion, we report the tunneling characteristic of graphene from the two-terminal devices after the breakdown. Negative differential conductance is also observed when a high voltage bias is applied across the graphene channel. The tunneling behavior could be attributed to the formation of non-uniform disordered graphene, which is created by the breakdown. We propose that the non-uniform disordered structure can introduce energy barriers in the graphene channel. This hypothesis is supported by the Raman images and the simulated results of the *I-V* characteristics from a one-dimensional single-square barrier. A 2 nm thick glassy carbon film, which is uniformly disordered, is compared and linear *I-V* characteristics of grassy carbon prove that the tunneling characteristic is a unique property of non-uniform disordered graphene. The observed memory switching effect up to a 100000% ON/OFF ratio may open up new possibilities for various graphene based applications and the tunneling effect paves a way to study graphene disordered characteristics such as defect magnetism or a weak localization in graphene.[18]

This work was supported by the Singapore Ministry of Education Academic Research Fund Tier 2 (MOE2008-T2-1-105).

References

[1] A. K. Geim, Science **324**, 1530 (2009).
[2] A. K. Geim and K. S. Novoselov, Nat. Mater. **6**, 183 (2007).
[3] Y. M. Lin, C. Dimitrakopoulos, K. A. Jenkins, D. B. Farmer, H. Y. Chiu, A. Grill, and P. Avouris, Science **327**, 662 (2010).
[4] S. Bae, H. Kim, Y. Lee, X. Xu, J.-S. Park, Y. Zheng, J. Balakrishnan, T. Lei, H. Ri Kim, Y. I. Song, Y.-J. Kim, K. S. Kim, B. Ozyilmaz, J.-H. Ahn, B. H. Hong, and S. Iijima, Nat. Nanotechnol. **5**, 574 (2010).
[5] B. Standley, W. Bao, H. Zhang, J. Bruck, C. N. Lau, and M. Bockrath, Nano Lett. **8**, 3345 (2008).
[6] I. Meric, M. Y. Han, A. F. Young, B. Ozyilmaz, P. Kim, and K. L. Shepard, Nat. Nanotechnol. **3**, 654 (2008).
[7] J. Yao, Z. Jin, L. Zhong, D. Natelson, and J. M. Tour, ACS Nano **3**, 4122 (2009).
[8] K. S. Novoselov, A. K. Geim, S. V. Morozov, D. Jiang, Y. Zhang, S. V. Dubonos, I. V. Grigorieva, and A. A. Firsov, Science **306**, 666 (2004).
[9] Y. J. Shin, Y. Wang, H. Huang, G. Kalon, A. T. S. Wee, Z. Shen, C. S. Bhatia, and H. Yang, Langmuir **26**, 3798 (2010).
[10] A. C. Ferrari, J. C. Meyer, V. Scardaci, C. Casiraghi, M. Lazzeri, F. Mauri, S. Piscanec, D. Jiang, K. S. Novoselov, S. Roth, and A. K. Geim, Phys. Rev. Lett. **97**, 187401 (2006).
[11] A. Barreiro, M. Lazzeri, J. Moser, F. Mauri, and A. Bachtold, Phys. Rev. Lett. **103**, 076601 (2009).
[12] E. Pop, D. Mann, J. Cao, Q. Wang, K. Goodson, and H. Dai, Phys. Rev. Lett. **95**, 155505 (2005).
[13] R. Murali, Y. Yang, K. Brenner, T. Beck, and J. D. Meindl, Appl. Phys. Lett. **94**, 243114 (2009).
[14] R. Waser and M. Aono, Nat. Mater. **6**, 833 (2007).
[15] H. L. Wang, X. R. Wang, X. L. Li, and H. J. Dai, Nano Res. **2**, 336 (2009).
[16] Y. Li, A. Sinitskii, and J. M. Tour, Nat. Mater. **7**, 966 (2008).
[17] P. G. Collins, M. Hersam, M. Arnold, R. Martel, and P. Avouris, Phys. Rev. Lett. **86**, 3128 (2001).
[18] O. V. Yazyev, Phys. Rev. Lett. **101**, 037203 (2008).

Figure Captions

Figure 1. (a) Raman spectra of single layer and multi-layer graphene. (b) Resistance vs. back gate voltage of a graphene sample. The inset in (b) shows the optical image of graphene with gold contacts (the scale bar is 8 µm). (c) *I-V* curve in the high bias range. The inset in (c) shows *I-V* curve in the low bias range. (d) Differential conductance versus bias voltage.

Figure 2. (a) *I-V* curve through an electrical breakdown. The insets in (a) show different Raman spectra measured at two different locations in the graphene channel after the breakdown. (b) *I-V* curve after breakdown. (c) Absolute value of current as a function of bias voltage in a logarithmic scale. The inset in (c) shows a scanning electron microscopy image of the graphene channel after the breakdown (the scale bar is 1 µm). (d) Optical image of the sample (top panel) and Raman images plotted by the intensity of *D* and 2*D* band (the scale bar is 4 µm). The dotted red line indicates the area of Raman imaging. The blue arrows show the direction of current flow through the graphene channel.

Figure 3. (a) Simulated *I-V* data. The insets show the energy diagrams of disordered graphene system without and with the bias voltage. (b) Repeated *I-V* curves after the breakdown. (c) *I-V* curve in the high bias range after the breakdown. The inset in (c) shows *I-V* curve in a low bias range after the breakdown. (d) *I-V* curve of a glassy carbon film. The inset in (d) shows the Raman spectra of glassy carbon.

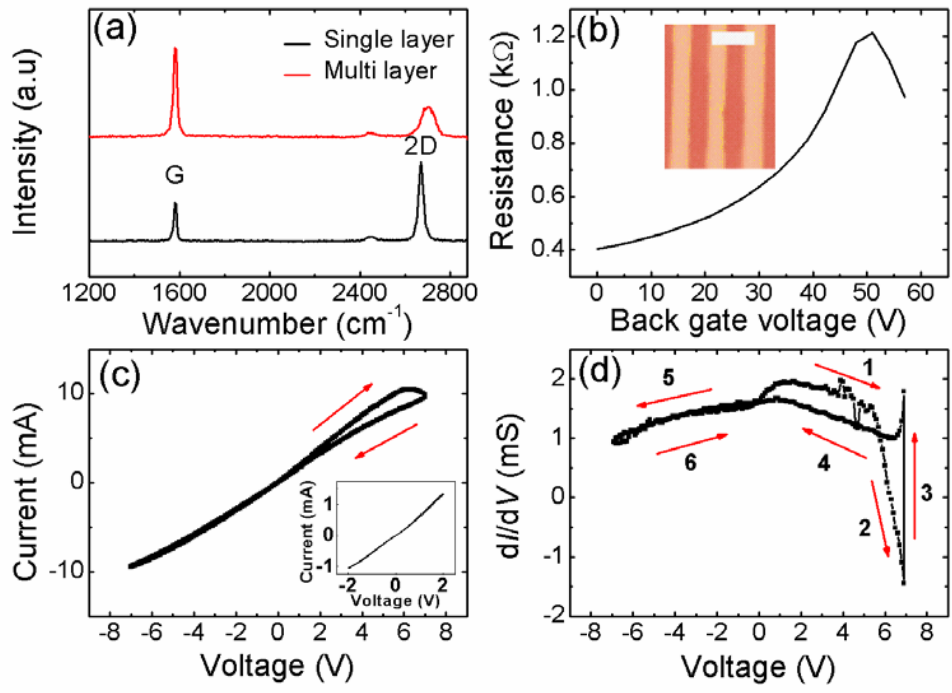

Figure 1.

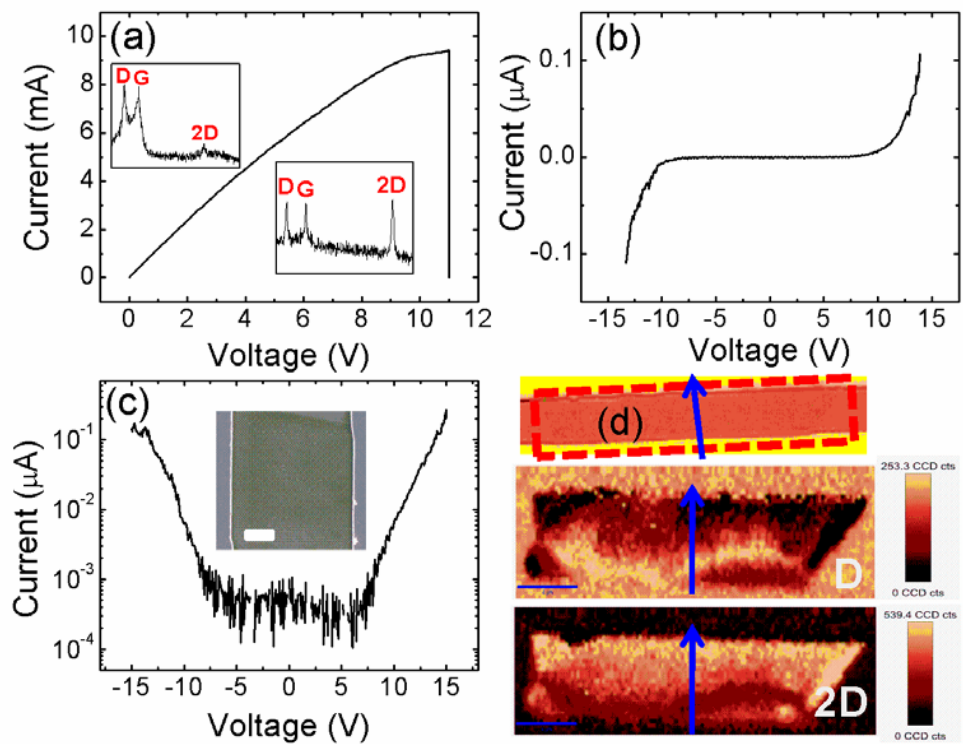

Figure 2.

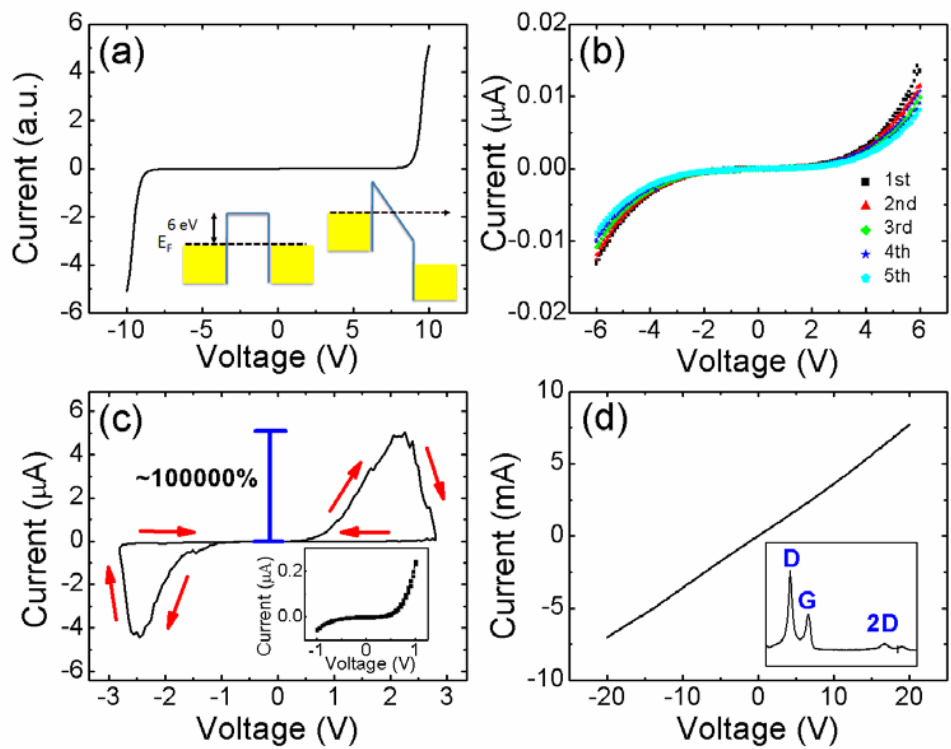

Figure 3.